\documentclass[10pt,conference]{IEEEtran}

\usepackage{amsmath,amssymb,amsfonts}
\usepackage{algorithmic}
\usepackage{graphicx}
\usepackage{textcomp}
\usepackage{flushend}
\usepackage{xcolor}
\usepackage{xspace}
\usepackage[numbers,sort]{natbib}

\usepackage{amsfonts} 
\usepackage{amsmath} 
\usepackage{booktabs} 
\usepackage{multirow} 
\usepackage{tablefootnote} 
\usepackage{enumitem} 
\usepackage{amsmath}  
\usepackage{subcaption}
\usepackage{scrextend} 
\usepackage{bbding} 
\usepackage[T1]{fontenc} 
\usepackage{enotez} 
\usepackage{tikz}
\usepackage[final]{pdfpages} 
\usepackage{ragged2e} 
\usepackage{array} 
\usepackage{adjustbox} 
\usepackage{threeparttable}
\usepackage{ltablex} 
\usepackage{url} 
\usepackage{framed} 

\usepackage{listings}
\usepackage{xcolor}

\usepackage{color}
\definecolor{dkgreen}{HTML}{72A98F}
\definecolor{mauve}{HTML}{834D93}
\definecolor{midnightblue}{HTML}{290DFD}

\newcommand{\adyen}{\emph{Adyen}\xspace}

\newcounter{observations}
\newcommand{\observation}[1]{\refstepcounter{observations} \vspace{2mm} \textbf{Observation \theobservations: #1}}
\setenotez{list-name=Wandb experiment links,mark-cs=\circledsuperscript, backref=true}
\newcommand*\circledsuperscript[1]{\textsuperscript{\scalebox{.9}{\tikz[baseline=(char.base)]{
            \node[shape=circle,draw,inner sep=2pt] (char) {#1};}}}}
\definecolor{pblue}{rgb}{0.13,0.13,1}
\definecolor{pgreen}{rgb}{0,0.5,0}
\definecolor{pred}{rgb}{0.9,0,0}
\definecolor{pgrey}{rgb}{0.46,0.45,0.48}
\begin{document}

\title{Learning Off-By-One Mistakes: An Empirical Study}

\author{\IEEEauthorblockN{Hendrig Sellik}
\IEEEauthorblockA{\textit{Delft University of Technology} \\
Delft, The Netherlands\\
sellikhendrig@gmail.com}
\and
\IEEEauthorblockN{Onno van Paridon}
\IEEEauthorblockA{\textit{Adyen N.V.} \\
Amsterdam, The Netherlands\\
onno.vanparidon@adyen.com}
\and
\IEEEauthorblockN{Georgios Gousios, Maurício Aniche}
\IEEEauthorblockA{\textit{Delft University of Technology} \\
Delft, The Netherlands\\
\{g.gousios,m.f.aniche\}@tudelft.nl}
}

\maketitle

\begin{abstract}

Mistakes in binary conditions are a source of error in many software systems. They happen when developers use, e.g., `$<$' or `$>$' instead of `$<=$' or `$>=$'. These boundary mistakes are hard to find and impose manual, labor-intensive work for software developers. 

While previous research has been proposing solutions to identify errors in boundary conditions, the problem remains open. In this paper, we explore the effectiveness of deep learning models in learning and predicting mistakes in boundary conditions. We train different models on approximately 1.6M examples with faults in different boundary conditions. We achieve a precision of 85\% and a recall of 84\% on a balanced dataset, but lower numbers in an imbalanced dataset. We also perform tests on 41 real-world boundary condition bugs found from GitHub, where the model shows only a modest performance. Finally, we test the model on a large-scale Java code base from \adyen, our industrial partner. The model reported 36 buggy methods, but none of them were confirmed by developers.
\end{abstract}


\section{Introduction}

\emph{Off-by-one} mistakes happen when developers do not correctly implement a boundary condition in the code. Such mistakes often occur when developers use `$>$' or `$<$' in cases where they should have used `$=>$' or `$<=$', or vice versa. 

Take the example of an off-by-one error in the Gson library\footnote{\url{https://github.com/google/gson/commit/161b4ba}}, which we illustrate in Figure~\ref{fig:gson}. The \texttt{toFind.length() < limit} condition is wrong. The fix changes the \texttt{<} operator by the \texttt{<=} operator.
Such mistakes are particularly difficult to find in source code. After all, the result of the program is not always obviously wrong, as it is ``merely off by one''. In most cases, the mistake will lead to an ``out of bounds'' situation, which will then result in an application crash. 

A large body of knowledge in the software testing field is dedicated to (manual) boundary testing techniques (e.g.,~\cite{jeng1994simplified, reid1997empirical, hoffman1999boundary, legeard2002automated, samuel2005boundary}).
However, manually inspecting code for off-by-one errors is time-consuming since determining which binary operator is the correct one is usually heavily context-dependent. The industry has been relying on static analysis tools, 
such as SpotBugs\footnote{\url{https://spotbugs.github.io}} or PVS-Studio\footnote{\url{https://www.viva64.com/en/pvs-studio/}}. SpotBugs promises to identify possible infinite loops, as well as array indices, offsets, lengths, and indexes that are out of bounds. PVS-Studio also tries to identify mistakes in conditional statements and indexes that are out of bounds in array manipulation. And while they can indeed find some of them, many of them go undetected. As we later show in this paper, none of the real-world off-by-one errors could be detected by the state-of-the-practice static analysis tools.

We conjecture that, for a tool to be able to precisely identify mistakes in boundary conditions, it should be able
to capture the overall context of the source code under analysis. Understanding the context of the source code has been traditionally a challenge for static analysis techniques. However, recent advances in machine and deep learning have shown that models can learn useful information from the syntactic and semantic information that exist in source code.
Tasks that were deemed not possible before, such as method naming~\cite{allamanis2014learning,allamanis2015suggesting,alon2019code2vec}, type inference~\cite{hellendoorn2018deep,allamanis2020typilus,pradel2019typewriter}, and bug finding~\cite{pradel2018deepbugs}, are now feasible.
The lack of reliable tools that detect off-by-one mistakes leaves an excellent opportunity for researchers to experiment with machine learning approaches. 

Inspired by the code2vec and code2seq models proposed by Alon et al.~\cite{alon2019code2vec,alon2018codeseq}, we trained several deep learning models on likely correct methods and their counterparts affected by off-by-one mistakes. 
The models are trained on over 1.6M examples, and the best results are obtained with the Code2Seq \cite{alon2018codeseq} model achieving 85\% precision and a recall of 84\% on a balanced testing set. 
However, our results also show that the model, when tested on a real-world dataset that consisted of 41 bugs in open-source systems, yields low performance (55\% precision and 46\% recall).

Finally, we tested the best models in one of our industrial partners. \adyen is one of the world's largest payment service providers allowing customers from over 150 countries to use over 250 payment methods including different internet bank transfers and point of sales solutions. The company is working in a highly regulated banking industry and combined with the high processing volumes there is little to no room for errors. Hence, \adyen uses the industry-standard best practices for early bug detection such as code reviews, unit testing, and static analysis. It is at \adyen's best interest to look into novel tools to prevent software defects finding their way into their large code base, preferring methods that scale and do not waste the most expensive resource of the company, the developers' time. 
Our results show that, while the model did not reveal any bugs per se, it pointed developers to code that they considered to deviate from their good practices.

\begin{figure}
\centering
\begin{lstlisting}[language=Java]
private boolean skipTo(String toFind) throws IOException {
  outer:
    for (; pos + toFind.length() < limit || fillBuffer(toFind.length()); pos++) {
      for (int c = 0; c < toFind.length(); c++) {
        if (buffer[pos + c] != toFind.charAt(c)) {
          continue outer;
  ...
}
\end{lstlisting}
\caption{A off-by-one error in the Gson library, fixed in commit \#161b4ba. The mistake is in the \texttt{toFind.length() < limit} condition; the fix changes the \texttt{<} by \texttt{<=}.}
\label{fig:gson}
\end{figure}

This paper expands our workshop paper, entitled ``OffSide: Learning to Identify Mistakes in Boundary Conditions''~\cite{briemoffside}.
The main contributions of this paper are:
\begin{enumerate}
\item An empirical study on the performance of different deep learning models, based on code2vec and code2seq, to detect off-by-one mistakes.
\item A quantitative and qualitative evaluation of deep off-by-one detection models in real-world open-source bugs and in a large-scale industrial system.
\end{enumerate}

\section{Related Work}

The use of static analysis tools is quite common among software development teams~(e.g.,~\cite{tomasdottir2018adoption,tomasdottir2017and}). These tools, however, rely on bug pattern detectors that are manually crafted and fine-tuned by static analysis experts. The vast amount of different bug patterns makes it very difficult to cover more than a fraction of them. 

Machine Learning for Software Engineering has seen rapid development in recent years inspired by the successful application in the Natural Language Processing field \cite{hindle2012naturalness}. It is applied in many tasks related to software code such as code translation (e.g.,~\cite{chen2018tree}), type inference (e.g.,~ \cite{hellendoorn2018deep,allamanis2020typilus}), code refactoring (e.g.,~ \cite{aniche2020effectiveness}) and, as we list below, bug identification. 

Pradel et al. \cite{pradel2018deepbugs} use a technique similar to Word2Vec \cite{mikolov2013distributed} to learn embeddings for JavaScript code tokens extracted from the AST. These embeddings are used to train two-layer feed-forward binary classification models to detect bugs. Each trained model focuses on a single bug type, and the authors test it on problems such as wrong binary operator, wrong operand in binary operation and swapped function arguments. These models do not use all the tokens from the code, but only those specific to the problem at hand. For example, the model that detects swapped function arguments only uses embeddings of the function name and arguments with a few other AST nodes as features. 

Allamanis et al. \cite{allamanis2017learning} use Gated Graph Neural Network \cite{li2016gated} to detect variable misuse bugs on a token level. As an input to the model, the authors use an AST graph of the source code and augment it with additional edges from the control flow graph.

Pascarella et al. \cite{pascarella2019fine} show that defective commits are often composed of both defective and non-defective files. They also train a model to predict defective files in a given commit. Habib et al. \cite{habib2019neural} create an embedding from methods using a one-hot encoding of tokens such as keywords (for, if, etc.), separators (;, (), etc.), identifiers (method, variable names and literals (values such as "abc" and 10). The embeddings for the first 50 tokens are then used to create a binary classification model. The oracle for training data is a state-of-the-art static analysis tool, and the results show that neural bug finding can be highly successful for some patterns, but fail at others.

Li et al. \cite{li2019improving} use method AST in combination with a global Program Dependency Graph and Data Flow Graph to determine whether the source code in a given method is buggy or not. The authors use Word2Vec to extract AST node embeddings with a combination of GRU Attention layer and Attention Convolutional Layer to build a representation of the method's body. Node2Vec \cite{grover2016node2vec} is used to create a distributed representation of the data flow graph of the file which the inspected method is in. The results are combined into a method vector which is used to make a softmax prediction.

Wang et al. \cite{wang2019learning} define bug prediction as a binary classification problem and train three different graph neural networks based on control flow graphs of Java code. They use a novel interval-based propagation mechanism to more efficiently generalize a Graph Neural Network (GNN). The resulting method embedding is fed into a feed-forward neural network to find null-pointer de-reference, array index out of bounds and class cast exceptions. For each bug type, a separate bug detector is trained.

\section{Approach}

In order to detect off-by-one errors in Java code, we aim to create a \textit{hypothesis function} that will calculate output based on the inputs generated from an example. More specifically, we train and compare different binary classification machine learning models to classify Java source code methods to one of the two possible output labels which are ''defective'' and ''non-defective''. If a method is considered as ''defective'', it is suffering from an off-by-one error, otherwise, it is deemed to be clear from errors.

These models are based on the Code2Vec~\cite{alon2019code2vec} and Code2Seq~\cite{alon2018codeseq} models, state-of-the-art deep learning models originally developed for generating method names and descriptions. The models use Abstract Syntax Tree paths of a method as features and create an embedding by combining them with the help of an attention mechanism. In addition, we also build a Random Forest baseline model based on source code tokens. 

We acquired the datasets necessary for the training of these models from the work of Alon et al.~\cite{alon2019code2vec} which results in an imbalanced dataset of 920K examples (1 to 10 ratio) and a balanced dataset of 1.6M examples when combined with our automatically mutated methods. 

We train on both imbalanced and balanced data to see the difference in performance. We then evaluate the accuracy of the model in 41 real-world open source off-by-one errors. In addition, we further train the models with data from a company project to fine-tune the model and find bugs from that project specifically. 

\begin{figure}
  \includegraphics[width=\columnwidth, keepaspectratio]{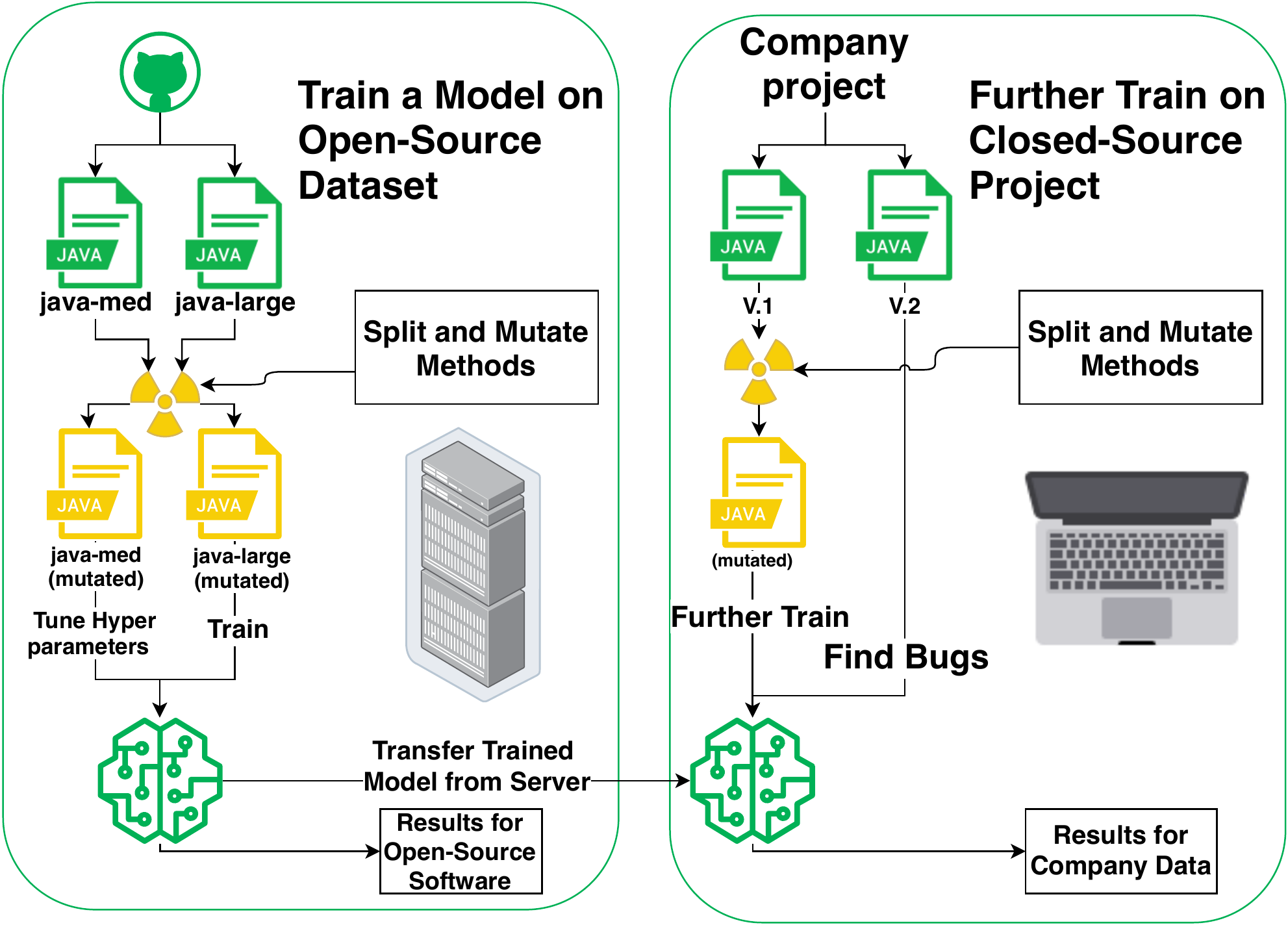}
  \caption{The flow of the research including data collection, mutation, training and testing.}
  \label{fig:research-flow}
\end{figure}

In Figure \ref{fig:research-flow}, we show the overall research method. In the following, we provide a more detailed description of the approach and research questions.

\subsection{Datasets}
\label{sec:datasets}

We used the java-large dataset provided by Alon et al. \cite{alon2018codeseq} for model training. We used \adyen's production Java code to further train and test the model with project-specific data. Finally, we used an additional real-world-bugs dataset to evaluate models on real-world bugs. A summary of the datasets can be seen in Table \ref{tab:datasets}.

\begin{table}
\centering
\caption{The different datasets used in this paper.}
\label{tab:datasets}
\begin{tabular}{lrrrr} 
\toprule
\textbf{Dataset} & \textbf{Train} & \textbf{Validation} & \textbf{Test} & \textbf{Total}\\
\midrule
large-balanced & 1,593,610 & 30,634 & 48,516 & \textbf{1,672,760} \\
large-imbalanced  & 876,485 & 16,849 & 26,684 & \textbf{920,018} \\
\adyen  & 11,032 & 690 & 3,148 & \textbf{14,870} \\
open-source bugs  & - & - & 82 & \textbf{82} \\
\bottomrule
\end{tabular}
\end{table}

\begin{enumerate}[leftmargin=*]


\item The \textit{java-large-balanced} dataset consists of 9,500 top-starred Java projects from GitHub created since January 2007. Out of those 9500 projects, 9000 were randomly selected for the training set, 250 for the validation set and the remaining 300 were used for the testing set. Originally, this dataset contained about 16M methods, but 836,380 were candidates for off-by-one errors (e.g. methods with loops and if conditions containing binary operator $<$, $<=$, $>$ or $>=$). After mutating the methods, the final balanced dataset consisted of 1,672,760 methods, 836,380 of them assumed to be correct and 836,380 assumed to be buggy. 

\item The additional imbalanced dataset \textit{java-large-imbalanced} was constructed to emulate more realistic data, where the majority of the code is not defective. A 10-to-1 ratio between non-defective and defective methods was chosen since it resulted in a high precision while having a reasonable recall. We empirically observed that upon increasing the ratio of non-defective methods even further, the model did not return possibly defective methods when running on \adyen's codebase. Meaning that if the ratio was higher than 10-to-1, the recall of the model became too low to use it. 

\item \textit{\adyen's code} is a repository containing the production Java code of the company. It consists of over 200,000 
methods out of which 7,435 contain a mutation candidate to produce an off-by-one error. After mutating the methods, this resulted in a balanced dataset containing 14,870 data points.

\item \textit{41 real-world bugs} in boundary conditions were used for manual evaluation. We extracted the bugs from the 500 most starred
GitHub Java projects. The analyzed projects were not part of the training and evaluation sets and thus are not seen by a model before testing. Using a Pydriller script \cite{spadini2018pydriller}, we extracted a list of candidate commits where authors made a change in a comparator (e.g., a ''\(>\)'' to ''\(=>\)''; ''\(<=\)'' to ''\(<\)'', etc.). This process returned a list of 1,571 candidate commits which were analyzed manually until 41 were confirmed to be off-by-one errors and added to the dataset. The manual analysis was stopped due to it being a very labor-intensive process.
\end{enumerate}

\subsection{Generating positive and negative instances}

In order to train a supervised binary classification model, we require defective examples. To get those, we modified the existing likely correct code to produce likely incorrect code. For each method, we found a list of possible mutation points and selected a random one. After this, we altered the selected binary expressions using JavaParser\footnote{JavaParser GitHub page \url{https://github.com/javaparser/javaparser/}} in a way to generate an off-by-one error. 

Due to changing only one of the expressions, the equivalent mutant problem does not exist\footnote{Equivalent mutant problem may exist, for example, if we mutate ``dead code''. However, we conjecture that this is a negligible problem and will not affect the results.} for the training examples, unless the original code was unreachable at the position of the mutation. It is also important to note that the datasets are split on a project level for the java-large dataset and on a sub-module level for \adyen's code. This means that the positive and the negative examples both end up in the same training, validation or test set. We did this to avoid evaluating model predictions on a code that only had one binary operator changed compared to the code that was used during training.

\subsection{Model Architecture}
The models we used in this work are based on the recent Code2Vec model \cite{alon2019code2vec} and its enhancement Code2Seq \cite{alon2018codeseq}, and a baseline model that makes use of random forest. We describe the models in more detail in the next sub-sections.

\subsubsection{Code2Vec}

The Code2Vec model created by Alon et al. \cite{alon2019code2vec} is a Neural Network model used to create embeddings from Java methods. These embeddings were used in the original work to predict method names.

The architecture of this model requires Java methods to be split into \textit{path contexts} based on the AST of the method. A path context is a random path between two nodes in the AST and consists of two terminal nodes \(x_{s}\), \(x_{t}\) and the path between those terminal nodes \(p_{j}\) which does not include the terminals. The embeddings for those terminal nodes and paths are learned during training and stored in two separate vocabularies. During training, these paths are concatenated to a single vector to create a \textit{context vector} \(c_{i}\) which has the length \(l\) of \(2 \cdot x_{s} + x_{p}\) where the length of \(x_{s}\) is equal to \(x_{t}\). 

The acquired context vectors \(c_{i}\)  for paths are passed through the same fully connected (dense) neural network layer (using the same weights). The network uses hyperbolic tangent activation function and dropout in order to generate a \textit{combined context vector} \(\tilde{c}_{i}\). The size of the dense layer allows controlling the size of the resulting context vector. 

The attention mechanism of the model works by using a global attention vector  \(a \in  \mathbb{R}^{h}\) which is initialized randomly and learned with the rest of the network. It is used to calculate attention weight \(a_i\) for each individual combined context vector \(\tilde{c}_{i}\).

It is possible that some methods are not with a large enough AST to generate the required number of context paths. For this dummy (masked) context paths are inputted to the model which get a value of zero for attention weight \(a_i\). This enables the model to use examples with the same shape.

During training, a tag vocabulary \(tags\_vocab \in  \mathbb{R}^{\left | Y \right | \times l}\)  is created where for each tag (label) \(y_{i} \in Y\) corresponds to an embedding of size \(l\). The tags are learned during training and in the task proposed by the authors, these represent method names.

A prediction for a new example is made by computing the normalized dot product between code vector \(v\) and each of the tag embeddings \(tags\_vocab_{i}\),
resulting in a probability for each tag \(y_{i}\). The higher the probability, the more likely the tag belongs to the method.

\subsubsection{Code2Seq}

The Code2Seq model created by Alon et al. \cite{alon2018codeseq} is a sequence-to-sequence model used to create embeddings from Java methods from which method descriptions are learned. The original work was used to generate sequences of natural language words to describe methods. 

Similarly to the Code2Vec model, the model works by generating random paths from the AST with a specified maximum length. Each path consists of 2 terminal tokens \(x_{s}\), \(x_{t}\) and the path between those terminal nodes \(p_{j}\) which, in Code2Seq, includes the terminal nodes \(p_{s}\), \(p_{t}\) \(\in p_{j}\), but not tokens. 

It is important to make a difference between terminal tokens and path nodes. The former are user-defined values, such as a number \textit{4} or variable called \textit{stringBuilder} while the latter come from a limited set of AST constructs such as NameExpr, BlockStmt, ReturnStmt. There are around 400 different node types that are predefined in the JavaParser implementation\footnote{JavaParser Node types \url{https://www.javadoc.io/doc/com.github.javaparser/javaparser-core/3.15.9/com/github/javaparser/ast/Node.html}}.

During training, the path nodes and the terminal tokens are encoded differently. Terminal tokens get partitioned into subtokens based on the camelCase notation, which is a standard coding convention in Java. For example, a terminal token \textit{stringBuilder} will be partitioned into \textit{string} and \textit{Builder}. The subtokens are turned into embeddings with a learned matrix \(E^{subtokens}\) and encoding is created for the entire token by adding the values for subtokens.

Paths of the AST are also split into nodes and each of the nodes corresponds to a value in a learned embedding matrix \(E^{nodes}\). These embeddings are fed into a bi-directional LSTM which final states result in a forward pass output \(\overset{\rightarrow}{h}\) and backward pass output \(\overset{\leftarrow}{h}\). These are concatenated to produce a path encoding.

As with the Code2Vec model, the encodings of the terminal nodes and the path are concatenated and the resulting encoding is an input to a dense layer with \textit{tanh} activation to create a \textit{combined context vector} \(\tilde{c}_{i}\).
Finally, to provide an initial state to the decoder, the representations of all \(n\) paths in a given method are averaged.

The decoder uses the initial state \(h_{0}\) to generate an output sequence while attending over all the combined context vectors \(\tilde{c}_{1}\text{, ..., }\tilde{c}_{n}\). The resulting output sequence represents a natural language description of the method. We adapted Code2Seq's sequence output to be $\{(0|1), \text{<eos>}\}$, i.e., a 1 or 0 token indicating the method being buggy or not buggy, and a token that ends the sequence.

The advantage of the Code2Seq model is in the way the context vectors \(\tilde{c}_{i}\) are created. In particular, due to splitting terminal nodes. The vocabulary of the terminal nodes yields greater flexibility towards different combinations of sub-token combinations. In addition, while Code2Vec embeds entire AST paths between terminals, the Code2Seq model only embeds sub-tokens. This results in fewer out-of-vocabulary examples and a far smaller model size. The model also has an order-of-magnitude fewer parameters compared to the Code2Vec model.

\subsubsection{Baseline Model}
We developed a baseline model to assess the performance of a simpler architecture. For this, we used a Random Forest model \cite{breiman2001random} and compared the performance with the same datasets. 

First, we tokenized the Java methods using leaf nodes of their respective ASTs. After this, all the tokens of the method were vectorized using the TF-IDF method. The vectorized tokens of one method comprised a training example for a Random Forest model. This model was then trained on all of the methods from java-large training set.


\subsection{Hyper-parameter optimization and model training}
\label{sec:hyperparameter-optimization}

For hyper-parameter optimization, we used Bayesian optimization \cite{snoek2012practical}. We selected model precision as the optimization parameter since high precision is required to obtain a usable defect prediction model. 
We used Bayesian optimization over other methods like random search or grid search because it enables us to generate a \textit{surrogate function} that is used to search the hyper-parameter space based on previous results and acts as intuition for parameter selection. This results in saving significantly more time because the actual model does not need to run as much due to wrong parameter ranges being discarded early in the process.

The hyper-parameters are optimized in the \textit{java-med-balanced}, another dataset made public by \citet{alon2019code2vec}. The dataset consists of 1,000 top-starred Java projects from GitHub. Out of those 1000 projects, 800 were randomly selected for the training set, 100 for validation set and the remaining 100 were used for the testing set. Originally, this dataset contained about 4M methods, but 170,295 were candidates for off-by-one errors (e.g. methods with loops and if conditions containing binary operator $<$, $<=$, $>$ or $>=$). This resulted in a balanced dataset of 340,590 methods, 170,295 of them assumed to be correct and 170,295 assumed to be buggy.

We ran optimization for four different scenarios. Two runs for the balanced java-medium dataset with Code2Vec model and Code2Seq models, respectively, and an additional two runs with the same models for imbalanced datasets. 
We used a machine with Intel(R) Xeon(R) CPU E5-2660 v3 processor running at 2.60GHz with a Tesla M60 graphics card.

Once the hyper-parameters were identified, we train the Code2Vec and Code2Seq models (as well as the baseline) using the balanced and imbalanced versions of the \textit{java-large} dataset, and perform further training with the source code of our industrial partner. We show the training time of the final models in Table~\ref{tab:training-time}.

\begin{table}
 \centering
 \caption{Model training times. B=Balanced dataset, I=Imbalanced dataset, E=number of epochs, T=time to train.}
 \label{tab:training-time}
\begin{tabular}{lrrrrrr}
\toprule
 & \multicolumn{2}{c}{Balanced} & \multicolumn{2}{c}{Imbalanced} & \multicolumn{2}{c}{\adyen} \\
 \midrule
Model & T & E & T & E & T & E \\
\midrule
Baseline & 5h33m & 1 & 1h59m & 1 & 48s & 1 \\
Code2Vec & 1d2h2m & 52 & 11h6m & 52 & 1h1m & 53 \\
Code2Seq & 3d18h18m & 14 & 2d14h 41m & 15 & 1h8m & 17 \\
\bottomrule
\end{tabular}
 
\end{table}

\subsection{Analysis}

We report the precision and recall of our models. 
Precision helps to evaluate the models' proneness to classify negative examples as positive. The latter is also known as false positive. This means that a model with high precision has a low false-positive rate and a model with low precision has a high false-positive rate. More formally, precision is the number of true positive (TP) predictions divided by the sum of true positive and false positive (FP) predictions.


For a bug detection model, low precision means a high number of false positives, making the developers spend their time checking a large number of errors reported by the model only to find very few predictions that are defective. This means that in this work, we prefer high precision for a bug-detection model. 

Monitoring precision alone is not enough since a model that is precise but only predicts few bugs per thousands of bugs is also not useful. Hence, recall is also measured. It measures the models' ability to find all the defective examples from the dataset. A recall of a model is low when it does not find many of the positive examples from the dataset and very high if it manages to find all of them. More formally, it is the number of true positive predictions divided by the sum of true positive and false negative predictions.


Ideally, a bug prediction model would find all of the bugs from the dataset and have a high recall score. However, deep learning networks usually do not achieve perfect precision and recall at the same time. For more difficult problems with a probabilistic model, there can be a trade-off. When increasing the threshold of the model confidence for the positive example, the recall will decline. For this reason, a sci-kit learn package was used to also make a precision-recall curve to observe the effect of the change in precision and recall upon changing the confidence of the model needed to classify an example as positive (defective).

\subsection{Reproducibility}

We provide all the source code (data collection, pre-processing, and machine learning models) in our online appendix~\cite{appendix}. The source code is also available in GitHub\footnote{\url{https://github.com/hsellik/thesis/tree/MSR-2021}}.

\section{Methodology}

The goal of this study is to measure the effectiveness of deep learning models in identifying off-by-one mistakes. To that aim, we propose three research questions:

\begin{itemize}
  \item \textbf{RQ\textsubscript{1}:	How do the models perform on a controlled dataset?}
  In order to obtain a vast quantity of data, we use a controlled dataset (see Section \ref{sec:datasets}). We train the models on the dataset and use metrics such as precision and recall to assess the performance. 

  \item \textbf{RQ\textsubscript{2}:	How well do the methods generalize to a dataset made of real-world bugs?}
  We mine a dataset of real-world off-by-one error bugs from GitHub issues of various open-source projects. Then we use a model to predict the error-proneness of a method before and after a fix. This will indicate how well the model works for real-world data. This evaluation will enable us to extract the precision metric and compare it to the one from RQ\textsubscript{1}.
  
  \item \textbf{RQ\textsubscript{3}:	Can the approach be used to find bugs from a large-scale industry project?}
  One useful application to an error-detection model is to analyze the existing project and notify of methods containing off-by-one errors. We make several runs where the model is firstly trained on a dataset with mutated code and then tested on real code to find such errors. In addition, we further train the model with a different version of the industry project to find errors in the future versions of the project. 
\end{itemize}

To answer $RQ_1$, we performed hyper-parameter optimization. After this, we selected the best hyper-parameter values and trained the model with randomly initialized parameters on the java-large dataset on the same machine as used for hyper-parameter optimization (see Section \ref{sec:hyperparameter-optimization}). We trained Code2Seq and Code2Vec models until there was no gain in precision for three epochs of training in the evaluation set. After this, we assessed the model on the testing set of java-large dataset.

The process was conducted for three different configurations of data. These were:
\begin{enumerate}
	\item BB - the training data was balanced (B) with the cross-validation and testing data also being balanced (B).
	\item BI - the training data was balanced (B) with the cross-validation and testing data being imbalanced (I).
	\item II - the training data was imbalanced (I) with the cross-validation and the testing data also being imbalanced (I).
\end{enumerate}

The data imbalance was inspired by the work of Habib et al. \cite{habib2019neural}, who reported that a bug detection model trained on a balanced dataset would have poor performance when testing on a more real-life scenario with imbalanced classes.

To answer $RQ_2$, we selected the best-performing model on the controlled java-large testing set (see Table \ref{tab:results-controlled-set}), which was the model based on the Code2Seq architecture. After this, the model was tested on the bugs and their fixes found from several real-world Java projects (open-source bugs dataset in Table \ref{tab:datasets}). 

Firstly, we tested the model on the correct code that was obtained from the GitHub diff after the change to see the classification performance on non-defective code. To test the model performance on defective code, we reverted the example to the state where the bug was present using the git version control system. After this, we recorded the model prediction on the defective method. 

In addition, as a way to compare our work with static analysis, we apply three popular static analyzers to the same set of defective and non-defective snippets: \emph{SpotBugs} (v.4.0.0-beta1), \emph{PVS-Studio} (v.7.04.34029), and the static analyzer integrated with \emph{IntelliJ IDEA} (v. 2019.2.3). 

To answer $RQ_3$, we trained the Code2Seq model only on the data generated from the company project, but the training did not start with randomly initialized weights. Instead, the process was started with the weights acquired after training on the java-large dataset (see Figure \ref{fig:research-flow}). 

We selected the Code2Seq based model because it had the best performance on the imbalanced testing set of the controlled \textit{java-large} set. We selected the performance on the imbalanced controlled set as a criterion since we assumed that the company project also contains more non-defective examples than defective ones. 

We used the pre-trained model because the company project alone did not contain enough data for the training process. Additionally, due to the architecture of the Code2Seq and Code2Vec models, the embeddings of terminal and AST node vocabularies did not receive additional updates during further training with company data. We trained the model until there was no gain in precision for three epochs on the validation set, and after this, we tested the model on the test set consisting of controlled \adyen data. 

We conducted additional checking on \adyen data by trying to find bugs in the most recent version of the project. More specifically, we updated the project to its most recent version using their git version control system, and without any modifications to their original code, we used the model to predict whether every Java method in their code base had a off-by-one mistake. We analyzed all bug predictions that were over a threshold of 0.8 to see if they contained bugs. The 0.8 threshold was defined after manual experimentation. We aimed at a set of methods that were large enough to bring us interesting conclusions, yet small enough to enable us to manually verify each of them.


\subsection{Threats to Validity}
In this section, we discuss the threats to the validity of this study and the actions we took to mitigate them.

\subsubsection{Internal validity} Our method performs mutations to generate faulty examples from likely correct code by editing one of the binary condition within the method. This means that while the correct examples represent a diverse set of methods from open-source projects, the likely incorrect methods may not represent a realistic distribution of real-world bugs. This affects the model that is being trained with those examples and also the testing results conducted on this data.

\subsubsection{External validity} While the work included a diverse set of open-source projects, the only closed-source project that was used during this study was \adyen's. Hence, the closed-source projects (in training and in validation) are under-represented in this study.

Moreover, we have only experimented with Java code snippets. While the approach seems to be generic enough to be applicable to any programming language, the results might vary given the particular way that developers write code in different communities. Therefore, more experimentation needs to be conducted before we can argue that the results generalize to any programming language.

\section{Results}

In the following sections, we present the results of our research questions.

\begin{table*}
\centering
\caption{Model results in controlled testing sets. BB stands for balanced training and testing set, II stands for imbalanced training set and testing set.}
\label{tab:results-controlled-set}

\begin{tabular}{l|rr|rr|rrrrrr} 
\toprule
 & \multicolumn{2}{|c|}{\textbf{Experiment BB}} & \multicolumn{2}{c|}{\textbf{Experiment BI}} & \multicolumn{6}{c}{\textbf{Experiment II}} \\ 
 & \multicolumn{2}{|c|}{Java-large} & \multicolumn{2}{c|}{Java-large} & \multicolumn{2}{c}{Java-large} & \multicolumn{2}{c}{\begin{tabular}[c]{@{}l@{}}\adyen data\\(cross-project)\end{tabular}} & \multicolumn{2}{c}{\begin{tabular}[c]{@{}c@{}}\adyen data\\(further trained)\end{tabular}} \\
\midrule

Model & Pr. \% & Re. \% & Pr. \% & Re. \% & Pr. \% & Re. \% & Pr. \% & Re. \% & Pr. \% & Re. \% \\
\midrule

Code2Seq & \textbf{85.23} & \textbf{84.82} & 36.08 & \textbf{84.86} & 83.04 & \textbf{42.34} & \textbf{71.15} & \textbf{24.66} & 66.66 & \textbf{30.66} \\
Code2Vec & 80.11 & 77.01 & 28.52 & 75.53 & 64.65 & 41 & 53.85 & 20.46 & 43.95 & 23.39 \\
Baseline & 50 & 49.08 & 8.99 & 49.18 & 17.86 & 0.15 & 0.0 & 0.0 & 9.25 & 0.92 \\
Offside~\cite{briemoffside}  & 80.9   & 75.6   &                          -      & -                               & -      & -                & -      & -                                                                              & -      & -                                                                                 \\

\bottomrule

\end{tabular}
\end{table*}

\subsection{RQ\textsubscript{1}: How do the models perform on a controlled dataset?}

\begin{figure*}
    \centering
  \begin{subfigure}{0.45\textwidth}
    \includegraphics[width=\textwidth]{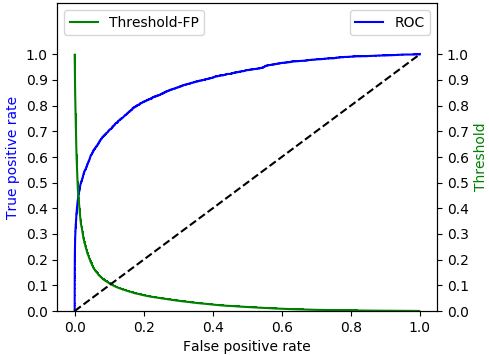}
    \caption{ROC curve. Area under curve 0.89.}
    \label{fig:code2seq-imbalanced-roc}
  \end{subfigure}
  \begin{subfigure}{0.45\textwidth}
    \includegraphics[width=\textwidth]{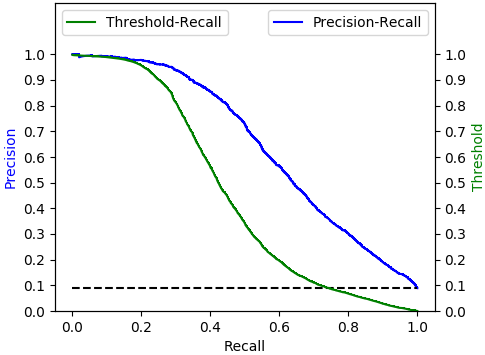}
    \caption{Precision-recall curve. Area under curve 0.65.}
    \label{fig:code2seq-imbalanced-precision-recall}
  \end{subfigure}
  \caption{ROC and precision-recall curves for the Experiment II with the java-large dataset.}
\end{figure*}

In Table \ref{tab:results-controlled-set}, we show the precision and recall of the different models. 
In Figures \ref{fig:code2seq-imbalanced-roc} and \ref{fig:code2seq-imbalanced-precision-recall}, we show the ROC curve and the precision-recall curve of the experiment with Code2Seq based model for the imbalanced java-large dataset. 

\observation{Models present high precision and recall when trained and tested with balanced data.} The results show that when training models on a balanced dataset with an equal amount of defective/non-defective code and then testing the same model on a balanced testing set, both Code2Vec and Code2Seq models achieve great precision and recall where the Code2Seq based model has better precision (85.23\% vs 80.11\%) and recall (84.82\% vs 77.01\%) compared to the Code2Vec based model. In addition, the balanced models' performance was compared to the one used in Offside~\cite{briemoffside}, our previous work exploring only the use of Code2Vec model, which was also tested on the identical java-large dataset using very similar preprocessing pipeline and training model (80.11\% vs 80.9\% precision and 77.01\% vs 75.6\% recall).

\observation{The metrics drop considerably when tested on an imbalanced dataset.} When simulating a more real-life scenario and creating an imbalance in the testing set with more non-defective methods, the recall of the models remained similar with recall increasing from 84.82 to 84.86 for the Code2Seq model and dropping from 77.01\% to 75.53\% for the Code2Vec model. However, the precision of the models reduced drastically with the Code2Seq model dropping from 85.23\% to 36.08\% and Code2Vec model from 80.11\% to 28.52\%.
The baseline model also drops in precision from 50\% to 8.99\% while keeping the same recall. 

\observation{The low precision can be mitigated by training on an imbalanced dataset, but at the cost of recall.} We trained Code2Seq and Code2Vec models on an imbalanced dataset and results show that the precision score for imbalanced data returned almost to the same level for the Code2Seq-based model (83.04\% vs 85.23\%), but remained lower for the Code2Vec-based model (64.65\% vs 80.11\%). However, the recall declined drastically from 84.82\% to 42.34\% for the Code2Seq model and from 77.01\% to 41.00\% for the Code2Vec model. 

When analysing the ROC curve (Figures~\ref{fig:code2seq-imbalanced-roc} and \ref{fig:code2seq-imbalanced-precision-recall}), the precision is $\approx$0.8 while recall remains $\approx$0.5 at a confidence threshold of 0.8. Moreover, it can also be seen that the model confidence is correlated where higher thresholds yield better precision but lower recall.

\vspace{2mm}
\underline{\textbf{RQ$_1$ summary:}} Both Code2Seq and Code2Vec based models present high accuracy on a balanced dataset. The numbers drop when we make use of imbalanced (i.e., more similar to the real-world) datasets.

\subsection{RQ\textsubscript{2}: How well do the methods generalize to a dataset made of real-world bugs?}

\begin{table}
\centering
\caption{Results of applying the Code2Seq model to 41 real-world off-by-one bugs and their corrected versions. B=balanced training set, I=imbalanced training set, Pr=Precision, Re=Recall, Thr=Threshold.}
\label{tab:rq2}
\resizebox{\linewidth}{!}{%
\begin{tabular}{lrrrrrrrr} 
\toprule
\textbf{Model} & \textbf{Thr} & \textbf{TP} & \textbf{TN} & \textbf{FP} & \textbf{FN} & \textbf{Pr} & \textbf{Re} & \textbf{F1} \\
\midrule
Code2Seq (B) & 0.5 & \textbf{19} & 26 & 15 & \textbf{22} & 55.88 & \textbf{46.34} & \textbf{50.67} \\
Code2Seq (B) & 0.8 & 10 & 33 & 8 & 31 & 55.56 & 24.39 & 33.9 \\
\midrule
Code2Seq (I) & 0.5 & 3 & \textbf{41} & \textbf{0} & 38 & \textbf{100} & 7.32 & 13.64 \\
Code2Seq (I) & 0.8 & 1 & \textbf{41} & \textbf{0} & 40 & \textbf{100} & 2.44 & 4.76 \\
\bottomrule
\end{tabular}}
\end{table}

The performance of the model on the 41 real-world boundary mistakes and their non-defective counterparts are presented in Table \ref{tab:rq2}.

\observation{The model can detect real-world bugs, but with a high false-positive rate.} Out of the 41 defective methods, 19 (46.34\%) were classified correctly and out of 41 correct methods, 26 (63.41\%) were classified correctly.
The precision and recall scores of 55.88 and 46.34 were achieved while evaluating the model on real-world bugs with the Code2Seq model trained on balanced data using a threshold of 0.5. Compared to the results from the java-large testing set with augmented methods, the results are significantly lower with precision and recall being 29.35 and 38.08 points lower respectively (see metrics for Code2Seq model with \textit{Experiment BB} in Table \ref{tab:results-controlled-set}).

\observation{The state-of-the-practice linter tools did not find any of the real-world bugs.}
As an interesting remark, none of the bugs was identified by any of the state-of-the-practice linting tools we experimented. This reinforces the need for approaches that identify such bugs (by means of static analysis or deep learning).

\vspace{2mm}
\underline{\textbf{RQ$_2$ summary:}} The model presents only reasonable performance on real-world off-by-one mistakes in open-source projects. Static analysis tools did not detect any bug.

\subsection{RQ\textsubscript{3}: Can the approach be used to find bugs from a large-scale industry project?}

We present the accuracy of the model in our industrial partner,
\adyen, also in Table \ref{tab:results-controlled-set}.

\observation{Models trained on open-source data show satisfactory results in the industry dataset.} Our empirical findings show that when a model is trained on an open-source dataset and then applied to the company project (following the same pipeline of mutating methods as to generate positive and negative instances), it will have good precision and recall scores with 71.15\% and 24.66\% for Code2Seq and somewhat lower 53.85\% and 20.46\% for Code2Vec model respectively. 

\observation{Further training on the \adyen project did not yield better results.} We hypothesized that training the model further on \adyen's code base would give a boost in precision and recall scores. The recall of the models improved by 6.0 percentage points for Code2Seq based model and 2.93 for Code2Vec based model. However, the precision of both models dropped by 4.49 percentage points for Code2Seq and 9.9 for Code2Vec. 

\observation{The model did not reveal any bugs, but 20\% of the reported methods were considered suspicious by the developers.} Running the model on a newer version of the repository reported 36 potential bugs with a confidence threshold over 0.8 (which we chose after experimenting with different thresholds and analyzing the number of suspicious methods the model returned that we considered feasible to manually investigate). While no bugs were found after manually analyzing all the reported snippets, we marked seven methods as suspicious. When we showed these methods to the developers, they agreed that, while not containing a bug per se, the seven methods deviate from good coding standards and should be refactored. More specifically, four methods had the for loop being initialized at a wrong index (i.e., the for loop was initialized with $i=1$, but inside the body, the code performed several $i-1$) and three snippets had hard-coded unusual constraints in the binary expression (i.e., $a>256$, where 256 is a specific business constraint). Interestingly, \citet{pradel2018deepbugs} also observed that models can sometimes point to pieces of code that are not buggy, but highly deviated from coding standards.

\observation{The model can potentially be useful at commit time, however, the number of false alarms is to be considered.} Fixing mistakes regarding good code practices for old pieces of software might not be considered worthwhile at large companies, given the possible unwanted changes to the behavior of the software. However, if such a system were to be employed during automated testing, the alerts might help developers to adhere to better practices. We observed the model pointing to relevant problems in 7 out of the 36 potential bugs (20\% of methods it identifies). While 20\% might be considered a low number, one might argue that inspecting 36 methods out of a code base that contains thousands of methods is not a costly operation and might be worth the effort. However, we still do not know the number of false negatives that the tool might give, as inspecting all the methods of the code base is unfeasible.

\vspace{2mm}
\underline{\textbf{RQ$_3$ summary:}} When tested on a large-scale industrial software system, the approach did not reveal any bugs per se, but pointed to code considered to deviate from good practices. 
\section{Future Work}

We see much room for improvement before these models can reliably identify off-by-one errors. In the following, we list the ones we believe to be most urgent:

\textbf{The need for more data points for the off-by-one problem.} In this paper, we leveraged the existing java-large dataset created by \citet{alon2018codeseq}. While the entire dataset was built on top of 9,500 GitHub projects and contained approximately 16M methods, only around 836k had binary conditions (e.g., methods with loops and ifs containing a $<$, $<=$, $>$ or $>=$). We augment this dataset to 1.6M by introducing the defective samples. Nevertheless, there is a big difference between 16M and 1.6M methods for training. As \citet{alon2019code2vec} argues: ``a simpler model with more data might perform better than a complex model with little data''. It should be part of any future work to devise a much larger dataset for the off-by-one problem and try the models we experiment here before proposing more complex models.

Moreover, our dataset contains fewer usages of >= or <= compared to usages of > or <, clearly representing the preferences of developers when coding such boundaries. These differences can lead to biased training and, as a result, we observed models tending to give false positive results in case of >= or <=. One way to mitigate the issue is to create a balanced dataset with a more equal distribution of binary operators, as well as the distribution of the places of their occurrence (if-conditions, for- and while-loops, ternary expressions, etc).

\textbf{The challenges of imbalanced data.} In this study, we explored the effects of balancing and imbalacing in the effectiveness of the model. However, the real imbalance of the problem in real life (i.e., the proportion between methods with off-by-one mistakes and methods without off-by-one mistakes) is unknown, although we strongly believe it to be imbalanced. Nevertheless, a 10:1 proportion enables us to have an initial understanding how models would handle such high imbalance. Our results show that it indeed negatively affects the performance of the model. Therefore, we suggest researchers to focus their attention on how to make these models better in face of imbalanced datasets. 

\textbf{The support for inter-procedural analysis.} Currently, our approach is only supporting the analysis of the AST of one method. However, the behaviour of a method, and the possibility of the bugs thereof, also depends on the contents of the other methods. For example, in recent research by Compton et al. \cite{compton2020embedding}, the embedding vectors from the Code2Vec model are concatenated to form an embedding for the entire class. Future work should explore whether class embeddings would perform better.

\textbf{Experimenting with different (and more recent) architectures.} In our work, we mainly looked at Code2Vec and Code2Seq models. We now see more recent models, such as the GREAT model proposed by \citet{hellendoorn2019global}, which uses transformers and also captures the data-flow of the code. We believe that data-flow information would enhance the performance of our models.

\textbf{Making use of Byte-Pair Encoding (BPE) techniques.} NLP models are often dependent on the vocabulary they are trained on. The out-of-vocabulary (OoV) problem also happens in this work. When testing the models trained on top of open-source data at \adyen, we had to replace unknown tokens by a generic \textit{UNK}. We conjecture that this may diminish the effectiveness of the models. We unfortunately did not measure the extent of how many times the \textit{UNK} token was used in our experiments. We plan to more precisely measure it in future replications of this work. In future work, we also plan to make use of techniques such as Byte-Pair Encoding (BPE)~\cite{gage1994new,sennrich2015neural}, which attempts to mitigate the impact of out-of-vocabulary tokens. We note that the use of BPE is becoming more and more common in software engineering models (e.g.,~\cite{karampatsis2020big, allamanis2020typilus, hellendoorn2019global}).

\textbf{A deeper understanding of the differences between our model and Pradel's and Sen's~\cite{pradel2018deepbugs} model.} The DeepBugs paper explores the effectiveness of deep learning models to a similar problem, which authors call ``Wrong Binary Operator''. The overall idea of their approach is similar to ours (in other words, their work also served as inspiration for this one): the negative instances (i.e., the buggy code) are generated through mutations in the positive code (i.e., non-buggy code), the code representation is a vector that is based on the embeddings of all the identifiers in the code, and the classification task is a feed-forward neural network that learns from the balanced set of positive and negative instances. Their results show an accuracy of 89\%-92\% in the controlled dataset (i.e., slightly higher than our results in RQ1), and a precision of 68\% in the manual analysis (i.e., higher than our results in RQ2). Interestingly, authors also observe that the model also reports non-buggy code which deviates from best practices (i.e., similar to our observations in RQ3). When designing this study, we did explicit compare the results to DeepBugs. The embeddings derived from code2vec/seq capture more information, and we conjectured that they would naturally supersede DeepBugs. 
We nevertheless see a few differences between both works: First, in their ``Wrong Binary Operator'' task, the mutation replaces the (correct) binary operator to any binary operator, e.g., a \texttt{i < length} can become a \texttt{i \% length}. In our case, we limit ourselves only to off-by-one mistakes, i.e., a correct \texttt{i < length} will always become a \texttt{i <= length}. We conjecture that this may increase the difficulty for the model to learn, as bugs are now slightly more subtle. Second, while the manual analysis conducted in the DeepBugs paper is performed on the testing set (which contains artificial bugs), our RQ2 explores the performance of the model in real-world bugs, i.e., bugs that were found and fixed by developers. This extra reality we bring to the experiment may be the reason for the lower performance. Finally, we assumed that more robust models such as code2vec and code2seq would better capture the intricacies of the off-by-one mistake. The model used in DeepBugs is simpler and yet as accurate as ours. More work is needed to understand the pros and cons of our model and how both works can be combined for the development of better and more accurate models.

\section{Conclusions}

Software development practices offer many techniques for detecting bugs at an early stage. However, these methods come with their challenges and are either too labor-intensive or leave a lot of room for improvement.
In this paper, we adapted recent state-of-the-art deep learning models to detect off-by-one errors in Java code, which are traditionally hard for static analysis tools due to their high dependency on context.

We concluded that the trained models, while effective in controlled datasets, still do not work well in real-world situations.
We see the use of deep learning models to identify off-by-one errors as promising. Nevertheless, there is still much room for improvement, and we hope that this paper helps researchers in paving the road for future studies in this direction.

\section*{Acknowledgments}

We thank Jón Arnar Briem, Jordi Smit, and Pavel Rapoport for their participation in the workshop version of this paper.

\renewcommand*{\bibfont}{\footnotesize}
\bibliographystyle{IEEEtranN}
\bibliography{paper}

\end{document}